\documentstyle{mn}

\input epsf

\begin{document}
\journal{Preprint astro-ph/9812125}
\title[Cosmic microwave background constraints on the epoch of 
reionization]{Cosmic microwave background constraints on the epoch of 
reionization}
\author[L.~M.~Griffiths, D.~Barbosa and A.~R.~Liddle]{Louise 
M.~Griffiths,$^1$ Domingos Barbosa$^{1,2}$ and Andrew R.~Liddle$^{1,2}$\\
$^1$Astronomy Centre, University of Sussex, Falmer, Brighton BN1 9QJ\\
$^2$Astrophysics Group, The Blackett Laboratory, Imperial College,
London SW7 2BZ (present address)}
\maketitle
\begin{abstract}
We use a compilation of cosmic microwave anisotropy data to constrain
the epoch of reionization in the Universe, as a function of
cosmological parameters. We consider spatially-flat cosmologies,
varying the matter density $\Omega_0$ (the flatness being restored by
a cosmological constant), the Hubble parameter $h$ and the spectral
index $n$ of the primordial power spectrum. Our results are quoted
both in terms of the maximum permitted optical depth to the
last-scattering surface, and in terms of the highest allowed
reionization redshift assuming instantaneous reionization. For
critical-density models, significantly-tilted power spectra are
excluded as they cannot fit the current data for any amount of
reionization, and even scale-invariant models must have an optical
depth to last scattering of below $0.3$. For the currently-favoured
low-density model with $\Omega_0 = 0.3$ and a cosmological constant,
the earliest reionization permitted to occur is at around redshift 35,
which roughly coincides with the highest estimate in the
literature. We provide general fitting functions for the maximum
permitted optical depth, as a function of cosmological parameters. We
do not consider the inclusion of tensor perturbations, but if present
they would strengthen the upper limits we quote.
\end{abstract}
\section{Introduction}

The absence of absorption by neutral hydrogen in quasar spectra, the
Gunn--Peterson effect (Gunn \& Peterson 1965; see also Steidel \&
Sargent 1987; Schneider et al.~1991; Webb 1992; Giallongo et
al.~1992,1994), tells us that the Universe must have reached a high
state of ionization by the redshift of the most distant known quasars,
around five.  Several mechanisms for reionization, which requires a
source of ultra-violet photons, have been discussed, and are extensively 
reviewed by Haiman \& Knox (1999).  In the two most
popular models, the sources are massive stars in the first generation
of galaxies, or early generations of quasars, and these models have
seen quite extensive investigation (Couchman \& Rees 1986; Shapiro \&
Giroux 1987; Donahue \& Shull 1991 amongst others).  Other
possibilities are that the reionization is caused by the release of
energy from a late-decaying particle, usually thought to be a neutrino
\cite{Sciama}, mechanical heating from supernovae driven winds
(Schwartz, Ostriker \& Yahil 1975; Ikeuchi 1981; Ostriker \& Cowie 1981) or
even by cosmic rays (Ginsburg \& Ozernoi 1965; Nath \& Bierman 1993) or by 
radiation from evaporating primordial black holes \cite{gib96}.

One of the most important consequences of reionization is the effect
on the anisotropies in the cosmic microwave background (CMB), again reviewed by 
Haiman \& Knox (1999).  Before
reionization, the microwave background photons have insufficient
energy to interact with the atoms, but after reionization they can
scatter from the liberated electrons.  This leads both to a distortion
of the blackbody spectrum and to a damping of the observed
anisotropies.  Typically, the number density of electrons after
reionization is low enough that only a fraction of the photons are
rescattered, so that some fraction of the original anisotropy is
preserved.

There has been continuing rapid progress in observations of microwave
background anisotropies, and it is now well established that there is
a rise in the spectrum around angular scales of one degree or so,
which is where one expects to see the first acoustic (or Doppler)
peak.  While the issue of whether or not there is an actual peak, with
the spectrum falling off again on yet smaller angular scales, remains
somewhat controversial, the existence of significant perturbations on
the degree scale already indicates that reionization cannot have
occurred extremely early, as that would have wiped out the anisotropy
signal. A detailed analysis of the current constraints on reionization
is our purpose in this paper. The earliest analysis of this type was
made by de Bernardis et al.~(1997), and more recently Adams et
al.~(1998) made a specific application to the decaying neutrino model.

We will work within the class of generalized cold dark matter (CDM)
models.  We consider a subset, where the dark matter is cold and the
spatial geometry flat, and we assume that the initial perturbations
are Gaussian and adiabatic, with a power-law form, as predicted by the
simplest models of inflation.  Qualitatively, the COBE DMR detections
(Smoot et al.~1992; Bennett et al.~1996) provided evidence supporting
this class of models, by showing that large angular scale fluctuations
have a spectrum close to a scale-invariant one.  Comparison with a
range of observations, including the galaxy cluster number density and
the galaxy power spectrum, have led to several different recipes
aiming at concordance, with the CDM models presently providing the
best framework for understanding the evolution of structure in the Universe.

We allow the possible existence of a cosmological constant, as
supported by recent observations of the magnitude--redshift relation
for Type Ia supernovae (Garnavich et al.~1998; Perlmutter et
al.~1998a,b; Riess et al.~1998; Schmidt et al.~1998).  We fix the baryon
density using nucleosynthesis (Schramm \& Turner 1998).  The
parameters we vary are therefore the matter density $\Omega_0$, the
Hubble parameter $h$ and the spectral index $n$ of the density
perturbations.  In this paper, we constrain the amount of reionization
as a function of these parameters, by carrying out a goodness-of-fit
test against a compilation of microwave anisotropy data.  We do not
consider the related question of finding the overall best-fitting
parameters, and in particular of whether the favoured parameter
regions are much altered by the inclusion of reionization, leaving
that to future work.

\section{Reionization and the optical depth}

\subsection{The optical depth}

First we briefly review the relation between reionization redshift and
the optical depth. The effect on the microwave background anisotropies
is mainly determined by the optical depth to scattering, and doesn't
depend too much on the exact reionization history; for illustration we
will imagine that the Universe makes a rapid transition from
neutrality to complete ionization.

If the electron number density is $n_{{\rm e}}$, and the Thomson
scattering cross-section $\sigma_{{\rm T}}$, then the optical depth
$\tau$ is defined as
\begin{equation}
\tau(t) = \sigma_{{\rm T}} \int_t^{t_0} n_{{\rm e}} (t) \, c \, dt \,,
\end{equation}
where $t_0$ is the present time. To obtain it as a function of
redshift, we proceed as follows. First we define the ionization
fraction $\chi(z) = n_{{\rm e}}/n_{{\rm p}}$, where $n_{{\rm p}}$ is
the proton density. Assuming a 24\% primordial helium fraction,
$n_{{\rm p}} = 0.88 \, n_{{\rm B}}$, where $n_{{\rm B}}$ is the baryon
number density, the present value of which is related to the baryon density
parameter by
\begin{equation}
\Omega_{{\rm B}} = \frac{8\pi G}{3 H_0^2} \, m_{{\rm p}} \, n_{{\rm B}} \,,
\end{equation}
with $m_{{\rm p}}$ being the proton mass and $H_0$ the Hubble parameter in
the usual units. For simplicity, we will assume that helium is fully
ionized as well as hydrogen; allowing for helium to be only singly
ionized is a small correction (as indeed is allowing for the neutrons
at all). Two useful relations are the redshift evolution of the
electron number density
\begin{equation}
n_{{\rm e}} \propto (1+z)^3 \,,
\end{equation}
and the time--redshift relation
\begin{equation}
\frac{dz}{dt} = -(1+z) H \,.
\end{equation}
They give
\begin{equation}
\tau(z) = n_{{\rm p},0} \, \sigma_{{\rm T}} \, c \int_0^z (1+z')^2 
        \, \frac{dz'}{H(z')} \, \chi(z') \,,
\end{equation}
where the `0' indicates the present value.

As long as the dominant matter is non-relativistic
\begin{equation}
\frac{\Omega(z) \, H^2(z)}{(1+z)^3} = {\rm const} = \Omega_0 H_0^2 \,,
\end{equation}
and we can write
\begin{equation}
\label{e:tauz}
\tau(z) = \tau^* \int_0^z \, (1+z')^{1/2} \, 
        \sqrt{\frac{\Omega(z')}{\Omega_0}} \, \chi(z') \, dz' \,,
\end{equation}
where 
\begin{equation}
\tau^* = \frac{3 H_0 \, \Omega_{{\rm B}} \, 
        \sigma_{{\rm T}}\, c}{8\pi G m_{{\rm p}}} \times 0.88 
        \simeq 0.061 \, \Omega_{{\rm B}} h \,,
\end{equation}
the last equality following simply by substituting in for all the
constants, with the usual definition of the Hubble constant $h$. A
useful equation for the redshift dependence of $\Omega$, again
assuming only non-relativistic matter and a flat spatial geometry, is
\begin{equation}
\Omega(z) = \Omega_0 \, \frac{(1+z)^3}{1-\Omega_0+(1+z)^3 \Omega_0} \,.
\end{equation}

\begin{figure}
\centering 
\leavevmode\epsfysize=6cm \epsfbox{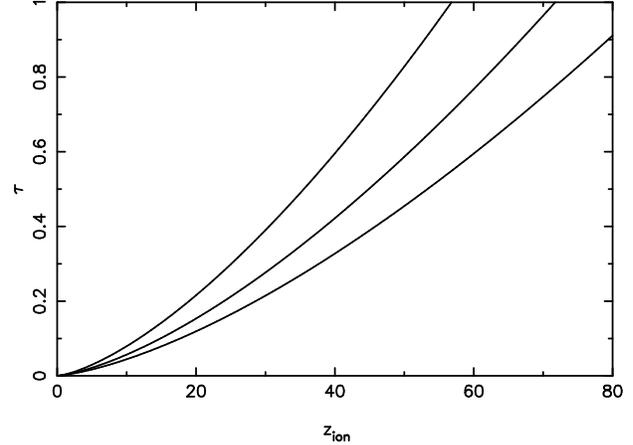}\\ 
\caption[tau_z]{\label{tau_z} Optical depth for instantaneous
reionization at redshift $z_{{\rm ion}}$. From top to bottom the
curves are $\Omega_0 = 0.3$, $0.6$ and $1$. We took $\Omega_{{\rm B}}
h^2 = 0.02$ and $h=0.65$.}
\end{figure}

For illustration we will assume instantaneous reionization at $z =
z_{{\rm ion}}$, so that $\chi(z) = 1$ for $z \leq z_{{\rm ion}}$ and
zero otherwise.  Equation~(\ref{e:tauz}) can then be integrated to
give
\begin{equation}
\label{e:tauz2}
\tau(z_{{\rm ion}}) = \frac{2\tau^* }{3\Omega_0} \, \left[ \left(1-\Omega_0
        + \Omega_0 (1+z_{{\rm ion}})^3 \right)^{1/2} - 1 \right] \,.
\end{equation}
Sample curves are shown in Figure~\ref{tau_z}. Inserting the latest
permitted reionization redshift, $z_{{\rm ion}}>5$ from the
Gunn--Peterson effect, implies only that $\tau$ exceeds a percent or
two for typical cosmological parameters. In order to give an optical
depth of unity, the epoch of reionization would be at
\begin{equation}
z \sim 100 \left( \frac{h\Omega_{{\rm B}}}{0.03} \right)^{-2/3}
        \Omega_0^{1/3} \,.
\end{equation}
Therefore, we should expect reionization to occur somewhere between 
$5 < z_{{\rm ion}} < 100$.

\subsection{Estimates of the reionization epoch}

Estimating the reionization redshift theoretically remains an
uncertain business. Structure formation in the CDM framework is
hierarchical, with the smallest gravitationally-bound systems forming
first and the bigger ones appearing later, by merging of the smaller
structures.  When the first fluctuations enter the non-linear growth
regime sometime after $z \sim 100$ (Peebles 1983), we expect the
appearance of the first bound objects and therefore, the possible
onset of reionization.  In most reionization models, the assumed
recipe is that baryons fall into the potential wells of the developing
structures in the cold dark matter, forming stars and quasars which
emit ultraviolet radiation.  When this radiation escapes the galaxies,
it will ionize and heat the intergalactic medium (IGM), and the usual
calculations aim to estimate when sufficient radiation is available to
complete the reionization. This is already a complex and uncertain calculation, 
made more so if one allows for inhomogeneities which can strongly affect the 
recombination rate (Carr, Bond \& Arnett 1984). Further, we should note that 
other heating
contributions are not currently excluded (Stebbins \& Silk 1986;
Tegmark, Silk \& Blanchard 1994; Tegmark \& Silk 1995) and may even be 
necessary.  Indeed, it has
been claimed from observations of the present UV background that it
may have been insufficient to reionize the IGM (Giroux \& Shapiro
1994), suggesting that collisional heating from supernovae-driven
winds or cosmic rays could also contribute to early reionization.

Density perturbation growth slows down with time, and structures in
low-density universes cease growing around
$1+z\sim1/\Omega_0$. Therefore, given the present observed matter power
spectrum, this implies that galaxies formed much earlier in low matter
density universes.  Consequently, reionization is expected to occur
earlier in low-density models and, given the bigger look-back time,
the optical depth will be larger.

The most extensive theoretical calculations, based on the
Press--Schechter mass function, tend to show that reionization
occurred after $z \sim 50$, and that a good guess for most models
would be $z_{{\rm ion}}\sim10-40$ (Fukugita \& Kawasaki 1994; Tegmark et 
al.~1994; Liddle \& Lyth 1995; Tegmark \& Silk 1995).
Low-density models are towards the top of this range and
critical-density ones towards the lower end (Liddle \& Lyth 1995).
These results have some corroboration from numerical simulations
(Haiman \& Loeb 1997).  Specifically, for $\Lambda$CDM models,
Ostriker \& Gnedin (1996) and Baltz, Gnedin \& Silk (1997) show that
reionization by population III stars should have sufficed to reionize
the IGM by $z \sim 20$, although recently Haiman (1998) suggested a
lower reionization redshift of around $z_{{\rm ion}}=9-13$ for a flat
low-density model.  If this is indeed the case, then besides the
determination of $z_{{\rm ion}}$ via damping of the CMB anisotropies
by CMB satellites MAP and Planck, the reionization redshift can be
measured {\it directly} from the spectra of individual sources with
the Next Generation Space Telescope (Haiman \& Loeb 1998) or with 21cm
``tomography'' with the Giant Meterwave Radio Telescope (Madau,
Meiksin \& Rees 1997).

In summary, the theoretical uncertainties in estimating the reionization 
redshift are large, and the plausible range stretches from just above the 
Gunn--Peterson limit of $z \simeq 5$ up to perhaps $40$. 

\subsection{Spectral distortions from reionization}

In the Thomson limit, where the incident photon energy in the electron
rest frame is much less than the electron rest mass--energy, the
blackbody form is preserved by scattering. However, the spectrum is
measured so accurately that one can hope to detect deviations
(Zel'dovich \& Sunyaev 1969). The best-known example is the
Sunyaev--Zel'dovich effect in clusters, which is detectable because of
the very high electron temperatures in clusters. The reionized
intergalactic medium is much cooler, but there is much more of it.
The amount of distortion of the CMB spectrum is defined through the
Compton $y$ parameter (Zel'dovich \& Sunyaev 1969; Stebbins \& Silk
1986; Bartlett \& Stebbins 1991):
\begin{equation}
y = \int \left( \frac{kT_{{\rm e}} - kT_{{\rm CMB}}}{m_{{\rm e}} c^2}
        \right) n_{{\rm e}} \sigma_{{\rm T}} c\;dt \,.
\end{equation}
At the epochs of interest, the CMB temperature is negligible compared
to the electron temperature, which we measure in units of $10^4$
Kelvin, denoted $T_4$. If the electron temperature is taken as
constant, this is the same integral as that giving the optical depth,
apart from the prefactor.

With the current limits on this distortion coming from the FIRAS
experiment, $y<1.5\times 10^{-5}$ (Fixsen et al.~1996), this implies,
for typical parameters,
\begin{equation}
z_{{\rm ion}} < 400 \, T_4^{-2/3} \left( \frac{h\Omega_{{\rm
        B}}}{0.03} \right)^{-2/3} \Omega_0^{1/3} \,.
\end{equation}
For the expected typical temperature evolution of the intergalactic
medium, it is hard to say much solely from the spectral distortions
about the reionization epoch, except that the Universe must have
undergone a neutral phase.  Stebbins \&
Silk (1986), Bartlett \& Stebbins (1991), Sethi \& Nath (1997) and more
recently Weller, Battye \& Albrecht (1998) show that almost no
reasonable reionized cosmological model violates current spectral
distortion constraints.  Therefore, the information coming from the
spatial damping of CMB anisotropies, rather than from spectral
distortions, is crucial to determine the history of the reionization
epoch, and from now we focus on the anisotropy power spectrum.

\section{The theoretical models}

As stated in the introduction, our aim is to constrain the epoch of
reionization for a range of spatially-flat cosmological models.  We
fix the baryon density at $\Omega_{{\rm B}} h^2 = 0.02$ from
nucleosynthesis \cite{ST98}.  This is at the high end of values
currently considered, which makes it a conservative choice because
decreasing the baryon density lowers the acoustic peak and hence
permits less reionization. We also do not consider tensor
perturbations, which contribute predominantly to the low
multipoles. As with the baryons, our constraints are conservative in
that they would strengthen if tensors were included, because including
them lowers the acoustic peaks relative to the low-$\ell$ plateau.

The three parameters we vary are
\begin{itemize}
\item $\Omega_0$ in the range $(0.2,1)$.
\item $h$ in the range $(0.5,0.8)$.
\item $n$ in the range $(0.8,1.2)$.
\end{itemize}
Our focus is directed towards obtaining upper limits on the amount of
reionization, though in some parts of parameter space there are lower
limits too.

The quantity to be compared with observation is the radiation angular
power spectrum $C_\ell$, which needs to be computed for each model.
The spherical harmonic index $\ell$ indicates roughly the angular size
probed, $\theta \sim 1/\ell$.  The power spectrum is readily
calculated using the {\sc cmbfast} program \cite{SZ}, which allows the
input of all the parameters we need.  However, exploring a
multi-dimensional parameter space is computationally quite intensive,
and rather than run {\sc cmbfast} for every choice of the optical
depth, it is more efficient and flexible to use an analytic
approximation to the effect of reionization.  This enables us to
quickly and accurately generate $C_\ell$ spectra for arbitrary amounts
of reionization.

The first step is to obtain spectra for the case with no reionization,
for each combination of our three parameters.  We take our parameters
on a discrete grid of dimensions $9 \times 7 \times 9$.  From these
spectra without reionization, we can generate spectra including
reionization using a version of the reionization damping envelope
technique of Hu \& White \shortcite{HW}.  This procedure readily
generates accurate enough spectra for comparison with the current
observational data, as we will show.  However, we do caution the
reader that this approach will not work once data of improved accuracy
becomes available.  Indeed, as shown by the Herculean 8-parameter
analysis of Tegmark (1999), the error bars on the cosmological
parameters coming from the CMB don't seem to change very much with the
addition of more parameters (see also Lineweaver 1998) and in the near
future the advent of better quality data will make them decrease,
introducing the necessity for a refined treatment of reionization.

\begin{figure}
\centering 
\leavevmode\epsfysize=7cm \epsfbox{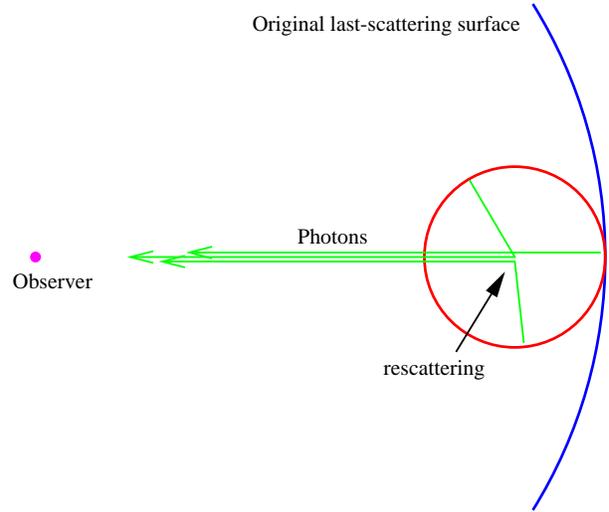}\\ 
\caption[rescatt]{\label{f:rescatt} We see a superposition of photons
from the original last-scattering surface, and those which
scattered. Of the latter, those which scattered once originated at decoupling on 
a smaller circle, whose size is given by the
time from decoupling to rescattering. Photons which scatter more than
once originate within this sphere.}
\end{figure}

The underlying physics is the following, illustrated in
Figure~\ref{f:rescatt}.  Given an optical depth $\tau$, the
probability that a photon we see originated at the original
last-scattering surface is $\exp(-\tau)$, the exponential accounting
for multiple scatterings. Those photons will still carry the original
anisotropy, which we will denote by $C_\ell^{{\rm int}}$. The remaining
fraction will have scattered at least once. Their contribution to the
anisotropy depends on scale. On large angular scales, they will have
rescattered within the same large region and will continue to share
the same temperature contrast; this is simply a statement that
causality prevents large-scale anisotropies being removed. On small
scales, however, the rescattered photons come from many different
regions with different small-scale temperature contrasts (the small
circle in Figure~\ref{f:rescatt}), and their anisotropy averages to
zero. Consequently we have two limiting behaviours for the observed
anisotropy $C_\ell^{{\rm obs}}$
\begin{eqnarray}
& C_\ell^{{\rm obs}} = C_\ell^{{\rm int}} & {\rm Small}\; \;  \ell \,; \\
& C_\ell^{{\rm obs}} = \exp(-2\tau) \, C_\ell^{{\rm int}} & {\rm Large} 
        \; \; \ell \,.
\end{eqnarray}
The factor 2 in the latter is because the power spectrum is the square
of the temperature anisotropy.

The reionization damping envelope \cite{HW} is a fitting function
which interpolates between these two regimes. For a given $\tau$, we
obtain the observed spectrum by
\begin{equation}
C_\ell^{{\rm obs}} = {\cal R}^2_\ell \, C_\ell^{{\rm int}} \,,
\end{equation}
where the reionization damping envelope ${\cal R}_\ell$ is given in
terms of the optical depth and a characteristic scale $\ell_{{\rm r}}$
by \cite{HW}
\begin{equation}
{\cal R}^2_\ell = \frac{1-\exp(-2\tau)}{1+c_1 x+c_2 x^2 +c_3 x^3 +c_4 x^4}+
        \exp(-2\tau) \,,
\end{equation}
with $x = \ell/(\ell_{{\rm r}} + 1)$ and $c_1 = -0.267$, $c_2 =
0.581$, $c_3 = -0.172$ and $c_4 = 0.0312$.

The characteristic scale comes from the angular scale subtended by the
horizon when the photons rescatter (i.e.~that subtended by the circle
in Figure~\ref{f:rescatt}). In order to obtain a highly accurate
result, Hu \& White \shortcite{HW} compute the characteristic scale
$\ell_{{\rm r}}$ via an integral which weights the horizon scale with
the optical depth, but for our purposes the simple formula
\begin{equation}
\ell_{{\rm r}} = \left( 1+z_{{\rm ion}} \right)^{1/2} 
        \left(1+0.084 \, \ln \Omega_0 \right)-1 \,, 
\end{equation}
gives sufficient accuracy, where $z_{{\rm ion}}$ is given by
rearranging equation~(\ref{e:tauz2}). This formula is a fit to the
angular size of the horizon at reionization (Hu \& White 1997, with a
sign error in their paper corrected).

\begin{figure}
\centering 
\leavevmode\epsfysize=6.15cm \epsfbox{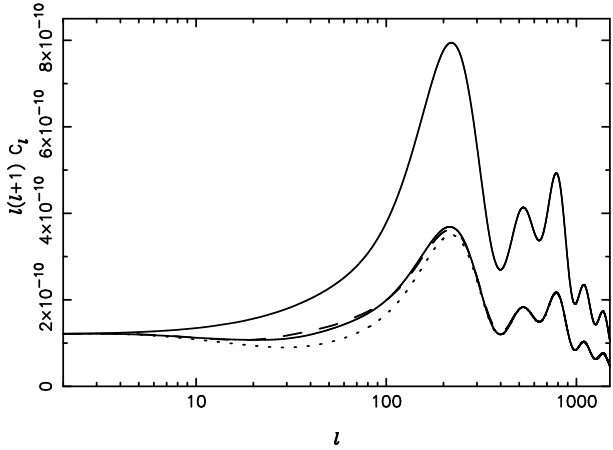}\\ 
\caption[cl_test]{\label{f:cl_test} Generating a $C_\ell$ curve
including reionization, illustrated for $n=1$, $h=0.5$, $\Omega_0=1$
and an optical depth $\tau = 0.4$. The top curve shows the spectrum
without reionization.  Applying the reionization damping envelope
generates the dotted line, and the correction for the new acoustic
peak, equation~(\ref{e:newpeak}), then gives the lower solid
line. This is to be compared with the exact result from {\sc cmbfast}
for this model, shown as the dashed line.}
\end{figure}

We are not quite finished yet, because while the reionization damping
envelope accounts for the loss of anisotropy due to scattering, it
does not allow for the generation of new anisotropies because of the
peculiar velocities of the rescattering electrons. These create a new,
but much less prominent, acoustic peak at smaller $\ell$ than the
original one. Because it is a minor feature, it can be modelled simply
using a Gaussian, the amplitude, width and location of which depend mildly on
the cosmology. The form we choose gives the reionized spectrum as
\begin{equation}
\label{e:newpeak}
C_\ell^{{\rm obs}} = \frac{{\cal R}^2_\ell \, C_\ell^{{\rm int}}}{1-f(\ell)} 
\,,
\end{equation}
where
\begin{eqnarray}
f(\ell) & = & A\exp \left( -\frac{1}{2\sigma^2} \, 
        \ln^2 \frac{\ell}{\ell_{{\rm max}}}  \right) \,, \\
A & = & \tau (\tau + 0.16) \,, \\
\sigma & = & 0.85 \,, \\
\ell_{{\rm max}} & = & 33 \Omega_0 + 21 h + 12.5 \tau \,.
\end{eqnarray}
The various numerical factors were fits from a comparison to {\sc
cmbfast} output in specific cases. This approach is easily accurate
enough given the current data, especially as the data are given in
$\delta T$ which corresponds to the square root of the $C_\ell$
curve. Figure~\ref{f:cl_test} shows an example compared to an exact
curve from {\sc cmbfast}.

\section{The observational data}

In recent years the detection of CMB anisotropies on different angular
scales has become commonplace.  At large scales, the COBE measurements
\cite{COBE,Ben96} constrained the amplitude of the spectrum with high
accuracy, and to some extent the slope.  Since then, a plethora of
ground-based and balloon-borne experiments probing medium and small
scales have followed, providing increasingly accurate measurements.
Although there is still quite a large scatter, there is very strong
evidence for the existence of an acoustic peak at $\ell$ of a few
hundred, as first claimed by Scott \& White (1994) and by Hancock \&
Rocha (1997), and therefore a limit on the amount of reionization
damping which is permitted.

\begin{figure}
\centering 
\leavevmode\epsfysize=6.15cm \epsfbox{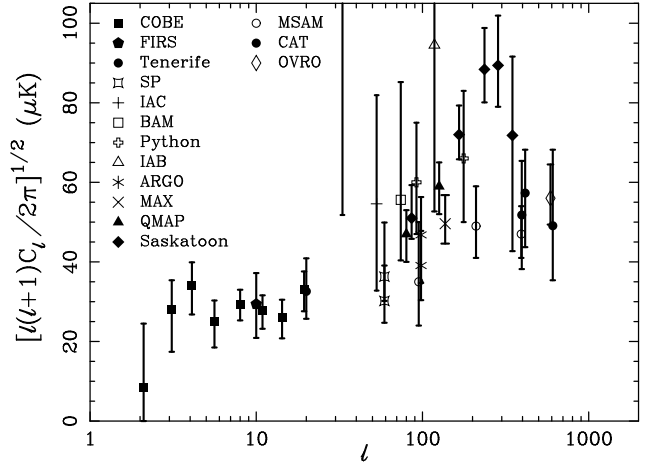}\\ 
\caption[Data_sample]{\label{f:Data_sample} The observed power
spectrum of CMB temperature fluctuations. Despite the scatter, there
is strong evidence of a rise to a peak at an $\ell$ of a few hundred.}
\end{figure}

\begin{table}
\begin{center}
\caption{\label{obsdat} The data used in this study, plotted in 
Figure~\ref{f:Data_sample}.}
\vspace{2pt}
\begin{tabular}{|l|l|r|c|} \hline         
Experiment & Reference &$\ell_{{\rm eff}}$ & 
$\delta T_{\ell_{{\rm eff}}}^{{\rm data}} \pm \sigma^{{\rm data}}(\mu$K)\\
\hline
COBE1   &1 &  2.1&$ 8.5_{-  8.5}^{+  16}$\\
COBE2   &1 &  3.1&$ 28.0_{-  10.6}^{+  7.4}$\\
COBE3   &1 &  4.1&$ 34.0_{-  7.2}^{+  5.9}$\\
COBE4   &1 &  5.6&$ 25.1_{-  6.6}^{+ 5.2}$\\
COBE5   &1 &  8  &$ 29.4_{-4.1}^{+ 3.6}$\\
COBE6   &1 & 10.9&$ 27.7_{- 4.5}^{+ 3.9}$\\
COBE7   &1 & 14.3&$ 26.1_{- 5.3}^{+ 4.4}$\\
COBE8   &1 & 19.4&$ 33_{- 5.4}^{+ 4.6}$\\
FIRS    &2 & 10&$ 29.4_{- 7.7}^{+  7.8}$\\
Tenerife &3 & 20&$ 32.6_{- 6.9}^{+ 8.3.}$\\
SP91    &4& 59&$ 30.2_{-  5.5}^{+  8.9}$\\
SP94    &4 & 59&$ 36.3_{-  6.1}^{+ 13.6}$\\
IAC1    &5&33 & $111.8_{- 60.0}^{+ 65.5}$\\
IAC2    &5&53 & $54.6_{- 21.8}^{+ 27.3}$\\ 
BAM     &6 & 74&$ 55.6_{- 15.2}^{+ 29.6}$\\
Pyth1   &7 & 92&$ 60.0_{- 13}^{+ 15}$\\
Pyth2   &7 &177&$ 66.0_{- 16}^{+ 17}$\\
IAB     &8 &118&$ 94.5_{- 41.8}^{+ 41.8}$\\
ARGO1   &$9^a$ & 98&$ 39.1_{-  8.7}^{+  8.7}$\\
ARGO2   &$9^b$ & 98&$ 46.8_{- 12.1}^{+  9.5}$\\
MAX     &10 &137&$ 46.9_{- 5}^{+ 7.2}$\\
QMAP1   &11 &80 &$ 49.0_{- 7}^{+6}$\\
QMAP2   &11 &126&$ 59.0_{- 7}^{+6}$\\
Sk1     &12 & 86&$ 51.0_{-  5.2}^{+  8.3}$\\
Sk2     &12 &166&$ 72.0_{-  6.2}^{+  7.3}$\\
Sk3     &12 &236&$ 88.4_{-  8.3}^{+ 10.4}$\\
Sk4     &12 &285&$ 89.4_{- 10.4}^{+ 12.5}$\\
Sk5     &12 &348&$ 71.8_{- 29.1}^{+ 19.8}$\\
MSAM    &13 &95&$ 35_{- 11.}^{+ 15.}$\\
MSAM    &13 &210&$ 49_{- 8.}^{+ 10.}$\\
MSAM    &13 &393&$ 47_{- 6}^{+ 7}$\\
CAT1    &$14^a$ &396&$ 50.8_{- 13.6}^{+ 13.6}$\\ 
CAT2    &$14^a$ &608&$ 49.1_{- 13.7}^{+ 19.1}$\\
CAT3    &$14^b$ &415&$ 57.3_{- 13.6}^{+ 10.9}$ \\
OVRO    &15 &589&$ 56.0_{- 6.6}^{+ 8.5}$\\
\hline
\end{tabular}
\end{center}
(1) Tegmark \& Hamilton 1997; (2) Ganga et al.~1994; (3) Guti\'errez et
al.~1997; Hancock et al.~1997 (binned); (4) Gundersen et al.~1995; (5)
Femenia et al. 1998; (6) Tucker et al.~1997; (7) Platt et al.~1997;
(8) Piccirillo \& Calisse 1993; ($9^a$) de Bernardis et al.~1994; ($9^b$) Masi 
et
al.~1996; (10) Tanaka et al.~1996 (binned); (11) de Oliveira-Costa et
al.~1998; (12) Netterfield et al.~1997; (13) Wilson et al.~1999; ($14^a$)
Scott et al.~1996 and Hancock \& Rocha 1997; ($14^b$) Baker et al.~1998
(15) Leitch et al.~1998.

\end{table}

Our data sample is shown in Figure~\ref{f:Data_sample}, and Table~1
lists the data points and indicates the sources from which they were
obtained. It is similar to the compilations described by Hancock \&
Rocha (1997) and Lineweaver et al.~(1997), and several researchers have 
up-to-date compilations available on the World Wide Web. We use
updated data and thus our sample includes:
\begin{itemize}
\item The 8 uncorrelated COBE DMR points from Tegmark \& Hamilton (1997).
\item The new calibration of the Saskatoon points (Leitch 1998); the shared 
calibration error of these points is small enough to be neglected.
\item The new updated QMAP results (de Oliveira-Costa et al.~1998).
\item The new OVRO Ring5M result (Leitch et al.~1998).
\end{itemize}

We use a $\chi^2$ goodness-of-fit analysis employing the data in
Table~1 along with the corresponding window functions, following the
method detailed by Lineweaver et al.~(1997). In brief, the window
functions describe how the anisotropies at different $\ell$ contribute
to the observed temperature anisotropies. For a given theoretical
model, they enable us to derive a prediction for the $\delta T$ which
that experiment would see, to be compared with the observations in
Table~1.

It has been noted that the use of the $\chi^2$ 
test can give a bias in parameter estimation in favour of permitting a lower 
power spectrum amplitude, as in reality there is a tail to high temperature 
fluctuations. Other methods have been proposed (Bond, Jaffe \& Knox 1998; 
Bartlett et al.~1999) which give good approximations to the true likelihood, 
though they require extra 
information on each experiment which is not yet readily available for the full 
compilation. We do not use these more sophisticated techniques here, but do 
note that as these methods are less forgiving of power spectra with too low an 
amplitude, the results from the $\chi^2$ analysis give conservative 
constraints on the optical depth.

\begin{figure}
\centering 
\leavevmode\epsfysize=6.2cm \epsfbox{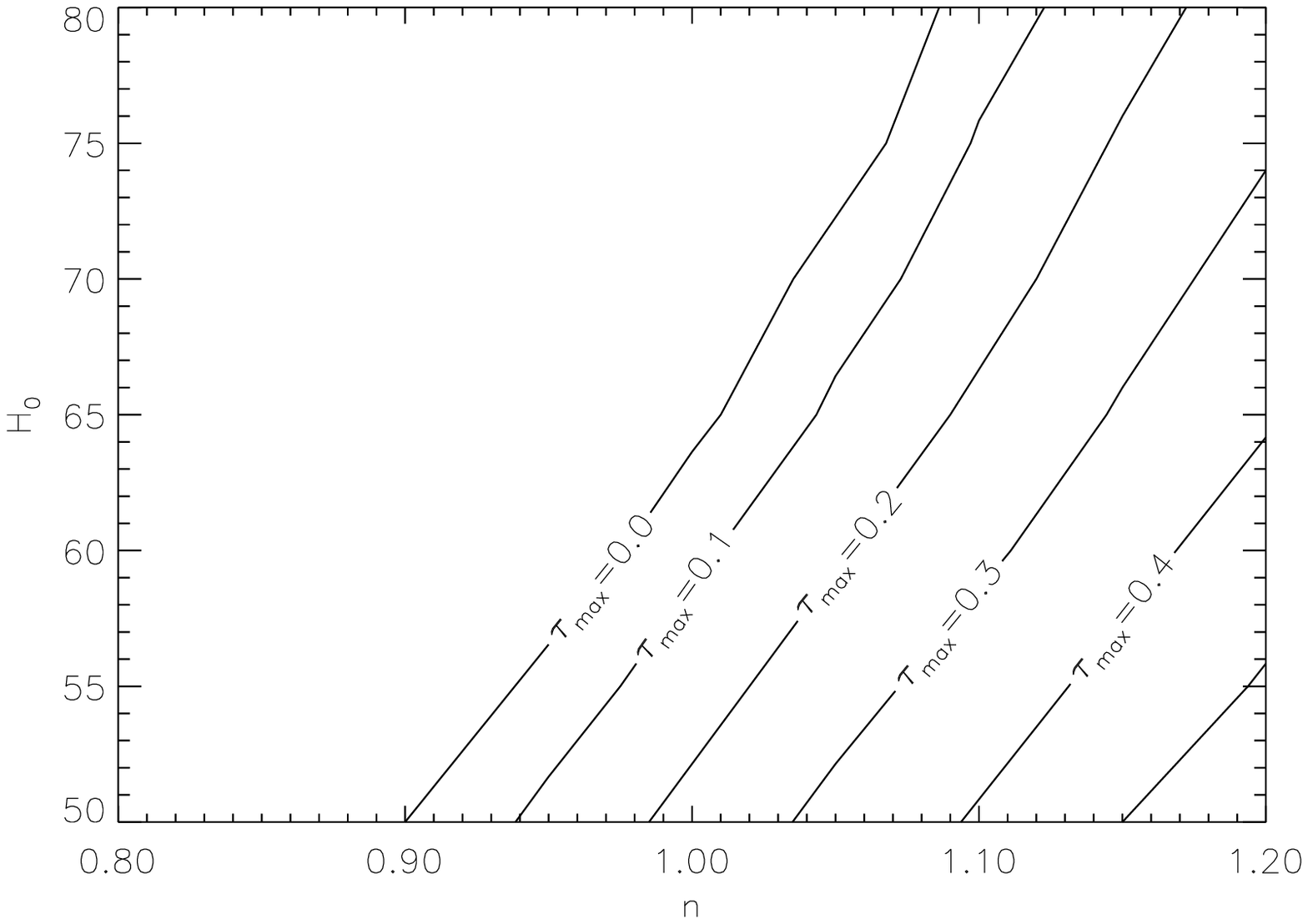}\\ 
\leavevmode\epsfysize=6.2cm \epsfbox{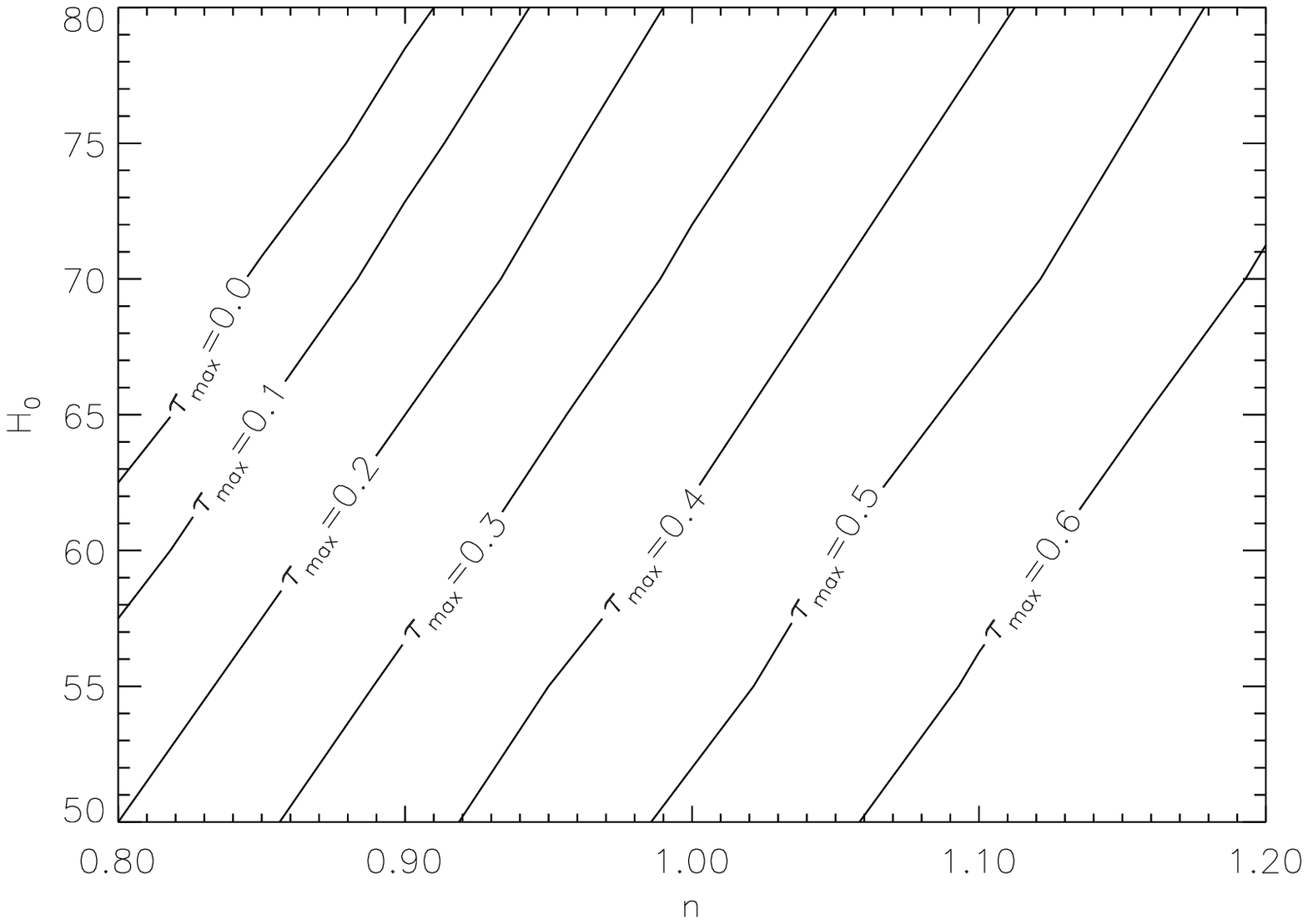}\\ 
\caption[omega]{\label{f:omega} Contours of the maximum permitted
optical depth, as a function of $n$ and $H_0$ at fixed $\Omega_0$. The
upper panel shows $\Omega_0 = 1$, the lower one $\Omega_0 =
0.3$. Regions to the left of the $\tau_{{\rm max}} = 0$ line are
excluded, as they do not allow a fit to the observational data for
{\em any} optical depth. The data are more constraining for  
$\Omega_0=1$, with an upper limit of $H_0 \la 65$ for a scalar invariant 
power spectrum.}
\end{figure}

\section{Constraints on the reionization epoch}

A model is specified by four parameters, $\Omega_0$, $h$, $n$ and
$\tau$.  There is an additional hidden parameter, which is the
normalization of the spectrum. We do not fix this by normalizing to
COBE alone, but rather seek the normalization which gives the best fit
to the entire data set. We then examine whether each model is a good
fit to the data.

There are $N_{{\rm data}} = 35$ data points. Because we are measuring
absolute goodness-of-fit on a model-by-model basis, with one hidden
parameter, the appropriate distribution for the $\chi^2$ statistic has
$N_{{\rm data}}-1$ degrees of freedom. Nothing further is to be
subtracted from this to allow for the main parameters, as they are not
being varied in the fit. To be specific, the question we are asking is
``If you are interested in particular values of $\Omega_0$, $h$, $n$
and $\tau$ for some reason other than the CMB data, will the predicted
CMB anisotropies be an adequate fit to the observations?''. To assess
whether a model is a good fit to the data, we need the confidence
levels of this distribution. These are
\begin{eqnarray}
\chi^2_{34} < 48.6 && \quad 95\% \; {\rm confidence \; level} \,; \\
\chi^2_{34} < 56.1 && \quad 99\% \; {\rm confidence \; level} \,.
\end{eqnarray}
Models which fail these criteria are rejected at the given level. We
will use the 95 per cent exclusion. Our main focus is on limiting
reionization, so for each choice of $\Omega_0$, $h$ and $n$, we are
interested in the largest value of $\tau$, $\tau_{{\rm max}}$, which
gives an acceptable fit.

Although we are not concerned with finding the
overall best-fitting parameters (which would require variation of
$\Omega_{{\rm B}}$ and ideally the inclusion of tensor perturbations,
as in Tegmark 1999), we note that the absolute best-fitting model in
our set, $\Omega_0 = 0.4$, $h = 0.6$, $n = 1.15$ and $\tau = 0.3$ has a
$\chi^2$ of 32, agreeing remarkably well with expectations
for a fit to 35 data points
with five adjustable parameters (the four mentioned plus
the amplitude). These $\chi^2$ values agree with other analyses of this type 
(Lineweaver 1998; Tegmark 1999), and the best-fit model has parameter values in 
excellent agreement with indications from other types of observation. 

\begin{figure}
\centering 
\leavevmode\epsfysize=6.2cm \epsfbox{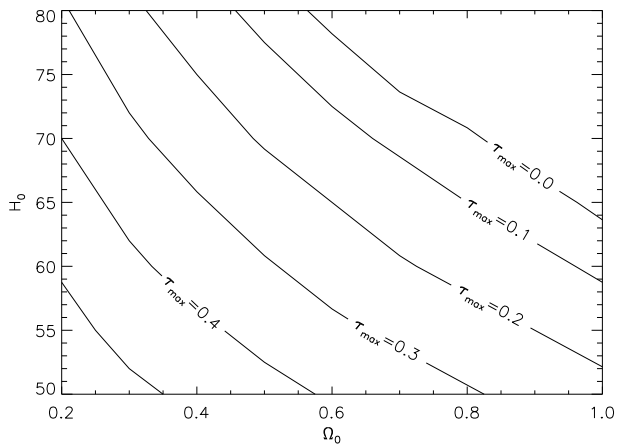}\\ 
\caption[n00]{\label{f:n00} Contours of the maximum permitted optical
depth, as a function of $\Omega_0$ and $H_0$ with $n=1$. }
\end{figure}

\begin{figure}
\centering 
\leavevmode\epsfysize=6.2cm \epsfbox{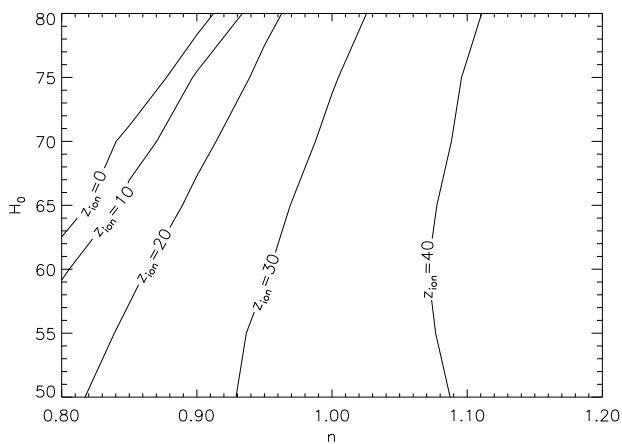}\\ 
\caption[zion03]{\label{f:zion03} Limits on the reionization redshift, for 
$\Omega_0 = 0.3$. Reionization must occur late than that indicated by the 
contour levels. This plot assumes instantaneous complete reionization. }
\end{figure}

The upper limits on the optical depth are shown in
Figures~\ref{f:omega} and \ref{f:n00}, for different slices across the
parameter space. For $\Omega_0 = 1$, quite a large amount of
otherwise-interesting parameter space is now excluded by the CMB data,
namely the region beyond the $\tau_{{\rm max}} = 0$ contour which will
not fit the data for any value of the optical depth. For the preferred
Hubble constant values of around $H_0 = 65 \, {\rm km} \, {\rm
s}^{-1}$, the lower limit on $n$ is now around $n = 1$, severely
constraining any attempts to salvage critical-density CDM models
through tilting the primordial spectrum. For critical density with
$n=1$, as commonly employed in mixed dark matter models, the optical
depth is constrained below $0.3$ or even $0.2$, depending on one's
preference for $H_0$ [note that the CMB anisotropies are hardly
altered by introduction of some hot dark matter in place of cold
\cite{dgs96}].

In the low-density case, the constraints on the optical depth are
weaker, because the first acoustic peak is predicted to be higher in the
absence of reionization. However, as there is a greater optical depth
out to a given redshift in low-density models, the constraints on the
actual reionization epoch prove to be quite similar. For $\Omega_0 =
0.3$, this is shown in Figure~\ref{f:zion03}, which was obtained from
the optical depth, assuming sufficiently-instantaneous reionization,
using equation~(\ref{e:tauz2}). We see that for the most
commonly discussed $n=1$ paradigm, the current limit on the
reionization redshift is around $z_{{\rm ion}} = 35$, which is just
about at the upper limit of the theoretically anticipated range
discussed in Section~2.2.  Future observations may well start to eat
into that range.

\begin{figure}
\centering 
\leavevmode\epsfysize=5.8cm  \epsfbox{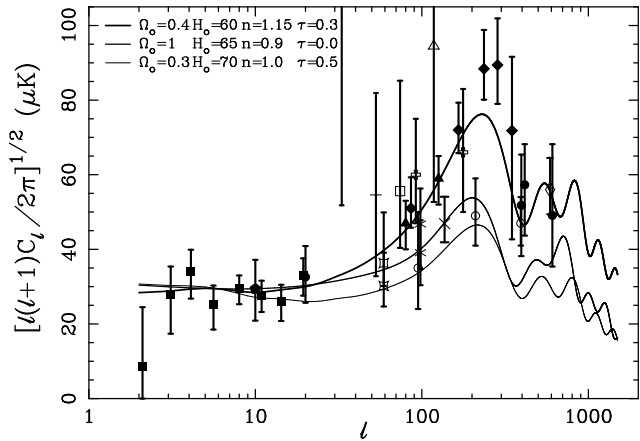}\\ 
\caption[model]{\label{f:model} The same data set of 
Figure~\ref{f:Data_sample}. The plotted curves show the best-fit 
model ($\Omega_0=0.4$ etc.) and two models that don't provide a 
reasonable fit. For the first model, the high optical depth compensates 
the gain of power at small scales caused by the tilt of the spectrum with 
$n>1$.}
\end{figure}

In Figure~\ref{f:model}, we show two typical models which fail to fit 
the data, as well as the absolute best-fitting model.

\begin{figure}
\centering 
\leavevmode\epsfysize=6cm \epsfbox{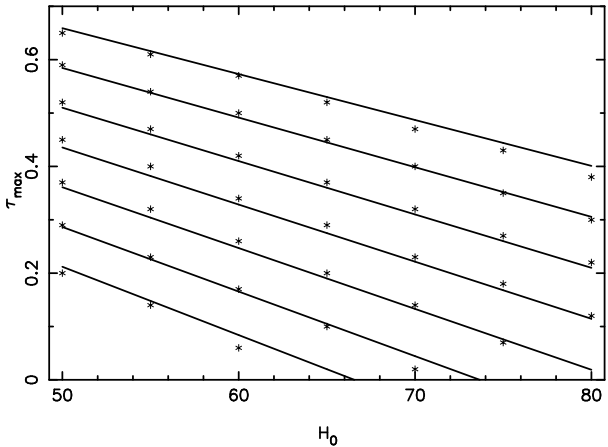}\\
\caption[fit_test]{\label{f:fit_test} An illustration of the fitting
function for $\Omega_0 = 0.3$. The lines show, from bottom upwards,
$n$ increasing from $0.8$ in steps of $0.05$. The points show the
exact results, for the $n$ of the line to which they are closest. The
worst error on $\tau_{{\rm max}}$ is around 0.02.}
\end{figure}

As well as describing the results graphically, it is useful to having
a fitting function for the maximum allowed optical depth. A good fit
for two particular $\Omega_0$ values is given by the formulae 
\begin{eqnarray}
\tau_{{\rm max}} \! \! \! & = \! \! \! & 0.03 - (2.9-1.5n) \, 
        (h-0.65) + 1.9 (n-1) \\ 
 && \hspace*{5.3cm} [\Omega_0 = 1] \,; \nonumber \\
\tau_{{\rm max}} \! \! \! & = \! \! \! & 0.36 - (2.4-1.4n) \, 
        (h-0.65) + 1.7 (n-1) \\
 && \hspace*{5.3cm} [\Omega_0 = 0.3] \,. \nonumber
\end{eqnarray}
The second of these is illustrated in Figure~\ref{f:fit_test}. For
general $\Omega_0$, a suitable interpolation between these is to
interpolate the three coefficients linearly in $\sqrt{\Omega_0}$
(e.g.~for the first coefficient take $0.76 - 0.73\,\sqrt{\Omega_0}$ and
so on).

There is no simple fitting function for the reionization redshift, but
an analytic fit is obtained by rearranging equation~(\ref{e:tauz2})
and putting in the fitting functions for $\tau_{{\rm max}}$.

\section{Summary}

We have developed an analytic method of generating the $C_\ell$
spectra in reionized models from models without reionization, and
confronted models with current observational data in order to place
upper limits on the optical depth caused by reionized electrons. We
stress that the constraint is best expressed on the optical depth, as
the main physical effect is that rescattered photons lose their short-scale
anisotropy and to a good approximation it doesn't matter where the
scattering took place. In general the optical depth is a function of
the complete reionization history, as well as the cosmological model,
but at least the first of these dependencies can be simplified if it
is assumed that reionization happens completely and fairly rapidly, in
which case the constraint can be re-expressed as an upper limit on the
reionization redshift.

We considered only a single value of the baryon density, at the high
end of the preferred range, and did not include tensor perturbations.
The second of these is definitely conservative, and the first likely
to be so, so our results can be regarded as rather safe upper limits.
However, these quantities would in general have to be included if one
undertakes the more ambitious task of trying to estimate best-fitting
parameters from the data, rather than delimiting the allowed region.
Several analyses have been carried out in recent years to use
available information to constrain the cosmological parameters, with
the majority neglecting the influence of reionization (Ganga, Ratra \&
Sugiyama 1996; White \& Silk 1996; Bond \& Jaffe 1997; 
Lineweaver et al.~1997; Bartlett et al.~1998; Hancock et
al.~1997; Lineweaver \& Barbosa 1998a, 1998b).  Most closely related
to this work are the papers of de Bernardis et al.~(1997) and more recently
Tegmark (1998), who investigated how reionization could affect
cosmological parameter determination.  Our results update and extend
the former paper, by employing more up-to-date data and exploring a
wider parameter space.  Neither of those papers aimed at providing
detailed constraints on the epoch of reionization, preferring instead
to find best-fitting parameters.  Although we have not made a serious
attempt at parameter estimation, we do concur with those papers that
the best-fitting models have a blue ($n>1$) spectrum and significant
reionization.

\section*{ACKNOWLEDGMENTS}

LMG was supported by the Nuffield Foundation under grant NUF-URB98, DB by the
European Union TMR programme and ARL in part by the Royal Society.  We thank
Joanne Baker, Kim Coble, Scott Dodelson, Marian Douspis, George Efstathiou, 
Morvan Le Dour, 
Charles Lineweaver, Avery Meiksin, Angelica de Oliveira-Costa, Rafael Rebolo, 
Antonio da Silva and Max Tegmark for helpful comments on this work, and the 
referee, Lloyd Knox, for an excellent report and further comments.  We 
acknowledge use of
the Starlink computer systems at the University of Sussex and at Imperial
College, and use of the {\sc cmbfast} code of Seljak \& Zaldarriaga (1996).



\bsp
\end{document}